\documentclass{ws-ijmpc}

\begin{document}

\title{Lattice Boltzmann kinetic modeling and simulation of thermal liquid-vapor system}
\author{YANBIAO GAN$^1$, AIGUO XU$^{2}$\thanks{Corresponding author. E-mail: Xu\_Aiguo@iapcm.ac.cn}, GUANGCAI ZHANG$^2$, JUNQI WANG$^3$, XIJUN YU$^2$,
 and YANG YANG$^4$}
\address{1 North China Institute of
Aerospace Engineering, Langfang 065000, P.R.China\\
2 Institute of Applied Physics and Computational Mathematics, P. O.
Box 8009-26, Beijing 100088, P.R.China\\
3 Shool of Renewable Resources, North China Electric Power University, Beijing  102206, P.R.China\\
4 China Petroleum Pipeline Material and Equipment Co., Ltd, Langfang 065000, P.R.China}
\catchline{}{}{}{}{}
\markboth{Y. Gan, A. Xu, G. Zhang, J. Wang, X. Yu and Y. Yang} {LB kinetic modeling and simulation of thermal liquid-vapor system}

\maketitle

\begin{abstract}
We present a highly efficient lattice Boltzmann (LB) kinetic model for
thermal liquid-vapor system. Three key components are as beow:
(i) a discrete velocity model by Kataoka \emph{et al.} [Phys. Rev. E \textbf{69},
035701(R)(2004)]; (ii) a forcing term $I_{i}$ aiming to describe the
interfacial stress and recover the van der Waals equation of state by
Gonnella \emph{et al.} [Phys. Rev. E \textbf{76}, 036703 (2007)]; and (iii)
a Windowed Fast Fourier Transform (WFFT) scheme and its inverse by our group
[Phys. Rev. E \textbf{84}, 046715 (2011)] for solving the spatial
derivatives, together with a second-order Runge-Kutta (RK) finite difference
scheme for solving the temporal derivative in the LB equation. The model is
verified and validated by well-known benchmark tests. The results recovered
from the present model are well consistent with previous ones[Phys. Rev. E \textbf{84}, 046715 (2011)]
  or theoretical
analysis. The usage of less discrete velocities, high-order RK algorithm and WFFT scheme with 16th-order in
precision makes the model more efficient by about $10$ times and more accurate than the
original one.
\end{abstract}

\catchline{}{}{}{}{}

\ccode{PACS Nos.: 47.11.-j, 47.20.Hw, 05.70.Ln}

\section{Introduction}

In the recent two decades the lattice Boltzmann (LB) method has been
becoming a promising simulation tool for various complex systems,\cite%
{succi,progphys} ranging from the low Mach number nearly incompressible flows%
\cite{LBGK} to high speed compressible flows under shocks,\cite%
{FOP_Review2012,EPL2013_LBGK} from simple fluids to multiphase and/or
multi-component fluids,\cite
{Yeomans,Sofonea-multiphase,PTXGL,PTGLS,PTGan} from phase
transition kinetics\cite{Yeomans,PTXGL,PTGLS,PTGan} to hydrodynamic
instabilities,\cite{Instability,FoP2013} etc. The LB methods in literature can be
roughly classified into two categories, new solvers of various partial
differential equations and kinetic models of various complex systems\cite{progphys}. In a
recent mini-review\cite{FOP_Review2012} it was pointed out that the LB
kinetic model can be used to investigate the macroscopic behavior of the
system due to its deviating from local thermodynamic equilibrium, including
cases where the Navier-Stokes models have already be there. This idea was
further specified in following works and considerable new physical insights
for various systems were obtained from a more fundamental level.\cite%
{progphys,EPL2013_LBGK,FoP2013}

Most early LB models for multiphase flows were for isothermal system and have been
successfully applied to a wide variety of flow problems, such as drop
collisions,\cite{POF-PRE-2005-Abraham} wetting,\cite
{SC-model--wetting-PRE-2006} phase separation and phase ordering,\cite%
{Yeomans,Sofonea-multiphase,PTXGL} etc. In recent years, some thermal LB
multiphase models\cite%
{PTGLS,PTGan,ZhangRY-PRE-IJHMT-PA-hybrid,Gonella-PRE-2010-hybrid,Markus-PRE-2011-DDF}
have appeared. Those models can be roughly classified into three categories, i.e., the
hybrid approach,\cite{ZhangRY-PRE-IJHMT-PA-hybrid,Gonella-PRE-2010-hybrid}
the double-distribution-function approach,\cite{Markus-PRE-2011-DDF} and the
multispeed-extra-force (MEF) approach.\cite{PTGLS,PTGan} In this paper we
focus on the MEF approach which has a rigorous theoretical background.
In the MEF approach, the one was developed by Gonnella, Lamura and Sofonea
(GLS)\cite{PTGLS} is typical since it considers all physical factors, e.g.,
viscous dissipation, interfacial tension, and compression work done by the
pressure, in both the momentum and energy equations.

A GLS-like model is formulated in two steps: firstly compose a thermal LB
(TLB) model for ideal gas system, then an extra force term $I_{i}$ is added
into the LB equation, where $I_{i}$ is responsible for
the interfacial stress and recovering the nonideal equation of state (EOS).
The GLS model was developed from the two-dimension 33-discrete-velocity
(D2V33) TLB model by Watari and Tsutahara,\cite{Watari-PRE-2003} and
utilized $I_{i}$ to reproduce, in the continuum limit, the full set of
thermohydrodynamic equations with the stress terms developed by
Onuki.\cite{Onuki-PRL-PRE-EPL} In a recent work\cite{PTGan}, to make better
the energy conservation and damp the spurious velocities to be negligible
small in practical simulations, the Windowed Fast Fourier Transform (WFFT) and its
inverse were introduced to calculate the spatial derivatives in the LB
equation. For convenience of description, we refer to this model as the
D2V33-FFT-TLB model.

The D2V33-FFT-TLB model is still subject to at least the following two
constraints that greatly hamper its wider applications: (i) limited density
ratio and temperature range; (ii) low computational efficiency. In the
present study, we address mainly the later restriction from both the
physical and computational point of views. In fact, computational efficiency
is of essential importance, especially for models proposed to
mimic the multiphase system. In a multiphase system without
\textquotedblleft fast factors" like shear, shock or strong convection, the
evolution of the system depends mainly on some \textquotedblleft slow
mechanisms", such as interface tension, viscous force, thermal diffusion,
etc. Generally, long lasting simulations are needed to establish the growth
properties.

\section{The model}

Here we present a highly efficient TLB model for van der Waals (VDW) fluids,
including the following three parts: (i) two-dimension
16-discrete-velocity (D2V16) LB model for ideal gas system\cite
{Kataoka-PRE-2004}, (ii) an appropriate interparticle force accounting for the
nonideal gas effects, and (iii) a higher-order windowed FFT (WFFT) scheme and
a 2nd Runge-Kutta finite difference scheme. For convenience of
description, we refer to this model as the D2V16-FFT-TLB model.

\subsection{D2V16 TLB model for ideal gas system}

Following the multispeed approach, Kataoka and Tsutahara proposed a TLB
model for compressible flows.\cite{Kataoka-PRE-2004} The model utilizes a discrete-velocity model
(DVM) that has only 16 discrete velocities.
The discrete equilibrium distribution function (DEDF) $f_{i}^{eq}$ is
calculated by a polynomial of the flow velocity $\mathbf{u}$ up to the fourth order.
Hydrodynamic quantities, such as density $\rho$, velocity $\mathbf{u}$, and temperature $T$ are calculated by
$\rho =\sum f_{i},\rho \mathbf{u}=\sum f_{i}\mathbf{v}_{i}$ and $\left( bT+%
\mathbf{u}^{2}\right) =\sum f_{i}\left( \mathbf{v}_{i}^{2}+\eta
_{i}^{2}\right)$, where $f_{i}$ is the discrete distribution function, $b$ represents the total degrees of freedom, $\eta_{i}$ is a free parameter introduced to
describe the $(b-2)$ extra degrees of freedom corresponding
to molecular rotation and/or vibration. Specifically, $\eta_{i}=2.5$ for $i=1,2,3,4$ and $\eta_{i}=0$ for $i=5,...,16$.

To the best of our knowledge, the D2V16 DVM is the one
with the least number of discrete velocities that can correctly recover the
hydrodynamic equations at the NS level. Physically, it is much faster than
other TLB models.

\subsection{The forcing term for multiphase flow system}

Although originally works only for ideal gas system, the multispeed model
can be extended to multiphase system using the extra force method. For
instance, GLS improved the D2V33 LB model by introducing an appropriate
force term, $I_{i}$, to describe the nonideal gas effects. Then the improved
equation reads,\cite{PTGLS}
\begin{equation}
\frac{\partial f_{i}}{\partial t}+\mathbf{v}_{i}\cdot \frac{\partial f_{i}}{%
\partial \mathbf{r}}=-\frac{1}{\tau }\left[ f_{i}-f_{i}^{eq}\right] +I_{i}%
\text{,}  \label{Iki_bgk}
\end{equation}%
where $I_{i}$ takes the following form:
\begin{equation}
I_{i}=-[A+\mathbf{B}\cdot (\mathbf{v}_{i}-\mathbf{u})+(C+C_{q})(\mathbf{v}%
_{i}-\mathbf{u})^{2}]f_{i}^{eq}\text{,}  \label{Iki}
\end{equation}%
with $A=-2(C+C_{q})T$, $\mathbf{B}=\frac{1}{\rho T}[\bm{\nabla}(P^{w}-\rho
T)+\bm{\nabla}\cdot \mathbf{\Lambda }-\bm{\nabla}(\zeta \bm{\nabla}\cdot
\mathbf{u})],C=\frac{1}{2\rho T^{2}}\{(P^{w}-\rho T)\bm{\nabla}\cdot \mathbf{%
u}+\mathbf{\Lambda \colon }\bm{\nabla}\mathbf{u}-\zeta (\bm{\nabla}\cdot
\mathbf{u})+\frac{9}{8}\rho ^{2}\bm{\nabla}\cdot \mathbf{u}+K[-\frac{1}{2}(%
\bm{\nabla}\rho \cdot \bm{\nabla}\rho )\bm{\nabla}\cdot \mathbf{u}-\rho %
\bm{\nabla}\rho \cdot \bm{\nabla}(\bm{\nabla}\cdot \mathbf{u})\mathbf{-}%
\bm{\nabla}\rho \mathbf{\cdot }\bm{\nabla}\mathbf{u\cdot }\bm{\nabla}\rho
]\} $, and $C_{q}=\frac{1}{\rho T^{2}}\bm{\nabla}\cdot \lbrack q\rho T%
\bm{\nabla}T]$.

In the continuum limit Eq. (1) can recover the following NSEs for VDW
fluids:
\begin{eqnarray}
\partial _{t}\rho +\bm{\nabla}\cdot (\rho \mathbf{u}) &=&0\text{,}  \nonumber
\\
\partial _{t}(\rho \mathbf{u})+\bm{\nabla}\cdot (\rho \mathbf{uu}+\mathbf{%
\Pi }-\bm{\sigma}) &=&0\text{,} \\
\partial _{t}e_{T}+\bm{\nabla}\cdot \lbrack e_{T}\mathbf{u+}(\mathbf{\Pi }-%
\bm{\sigma})\cdot \mathbf{u}-\kappa _{T}\bm{\nabla}T] &=&0\text{,}  \nonumber
\end{eqnarray}%
where $\mathbf{\Pi }=P^{w}\mathbf{I}+\mathbf{\Lambda }$ is the reversible
part of stress, comprising the VDW EOS $P^{w}=3\rho T/(3-\rho )-9/8\rho ^{2}$
and the nonideal gas interaction term $\mathbf{\ \Lambda }=M\bm{\nabla}\rho %
\bm{\nabla}\rho -M(\rho \nabla ^{2}\rho +\left\vert \bm{\nabla}\rho
\right\vert ^{2}/2)\mathbf{I}-[\rho T\bm{\nabla}\rho \cdot \bm{\nabla}(M/T)]%
\mathbf{I}$, with $M=K+HT$, $K$ is the surface tension coefficient and $H$ a
constant, $\mathbf{I}$ the unit tensor. $\bm{\sigma}=\eta \lbrack \bm{\nabla}%
\mathbf{u}+(\bm{\nabla}\mathbf{\mathbf{u}})^{T}-(\bm{\nabla}\cdot \mathbf{u})%
\mathbf{I}]+\zeta (\bm{\nabla}\cdot \mathbf{u})\mathbf{I}$ is the usual
viscous stress tensor with the shear and bulk viscosities $\eta $ and $\zeta
$. $e_{T}=\rho T-9\rho ^{2}/8+K\left\vert \bm{\nabla}\rho
\right\vert ^{2}/2+\rho u^{2}/2$ is the total energy density.

By applying Chapman-Enskog procedure, it is interesting to find that the
incorporation of the force term in the LB equation makes no additional
constrains on the requirements or the formulation process of the DEDF. In
other words, if the DEDF satisfies the seven kinetic moments, i.e., Eqs.
(1)-(7) as listed in Ref. [16], then the \textquotedblleft original LB
model + extra forcing term" is suitable for describing the multiphase flow
system, without considering the specific formulation of the DEDF and the
configuration of DVM. Consequently, the forcing term $I_{i}$ can be directly
applied to other LB models or DVMs under the same framework, e.g., the
aforementioned D2V16 model\cite{Kataoka-PRE-2004} or the D2V61 model\cite%
{Xu-EPL-2005} so as to reduce the computational load or improve the
isotropy. Here we choose the D2V16 TLB model to replace the D2V33 TLB model
so as to increase the computational efficiency.

\subsection{Numerical schemes for spatial and temporal derivatives}
From Ref. [8], it is well known that $k_{x}$ decides the accuracy of the FFT scheme.
Here we further develop the WFFT scheme through applying the $k_{x}$ with 16th order in
accuracy, i.e., $k_{x}$ takes the following form:\cite
{PTGan}
\begin{equation}
k_{x}=\frac{\arcsin\varphi}{\Delta x/2}
\simeq \frac{1}{\Delta x/2}(\varphi +\frac{1}{6}\varphi ^{3}+\frac{3}{40%
}\varphi ^{5}+\frac{5}{112}\varphi ^{7}+\frac{35}{1152}\varphi ^{9}+\frac{63%
}{2816}\varphi ^{11}+\frac{231}{13312}\varphi ^{13}+\frac{143}{10240}\varphi
^{15})\text{.}
\end{equation}%
where $\varphi =\sin (k_{x}\Delta x/2)$, and $k_{y}$ can be obtained in a similar way. Some simple derivations demonstrate
that the FFT approach with $k_{x}$ has a 16th-order accuracy in space.\cite
{PTGan} The time evolution of $f_{i}$ is derived by the 2nd finite
difference scheme\cite{2nd} $f_{i}^{n+1/2}=f_{i}^{n}-[{{{\mathbf{v}}}}_{i}\cdot {\frac{%
\partial }{\partial \mathbf{r}}}f_{i}^{n}{-\frac{1}{\tau }(}%
f_{i}^{n}-f_{i}^{eq,n})+I_{i}^{n}]\frac{\Delta t}{2}$ and $%
f_{i}^{n+1}=f_{i}^{n}-[{{\mathbf{v}}}_{i}\cdot \frac{\partial }{\partial
\mathbf{r}}f_{i}^{n+1/2}-\frac{1}{\tau }%
(f_{i}^{n+1/2}-f_{i}^{eq,n+1/2})+I_{i}^{n+1/2}]\Delta t$, where the
superscripts $n$, $n+1$ indicate the consecutive two iteration steps.

\section{Simulation results and physical analysis}

In this section, several numerical investigations are conducted to validate
the physical properties of the D2V16-FFT-TLB model.

\subsection{Liquid-vapor coexistence curve}

To check if the D2V16-FFT-TLB model can reproduce the correct equilibrium thermodynamics
of the multiphase system, simulations about the
liquid-vapor coexistence curve at various temperatures are performed.
Simulations are carried out over a $512\times 2$ domain with Periodic
Boundary Conditions (PBCs) in both the $x$- and $y$-directions.
The initial conditions are set as $(\rho, T, u_{x}, u_{y})|_{L}=(\rho _{v}, T_{0},
0.0, 0.0)$ if $x\leq N_{x}/4$, $(\rho, T, u_{x}, u_{y})|_{M}=(\rho _{l}, T_{0},
0.0, 0.0)$ if $N_{x}/4<x\leq 3N_{x}/4$; and $(\rho, T, u_{x}, u_{y})|_{R}=(\rho _{v}, 0.9975,
0.0, 0.0)$ if $3N_{x}/4<x$, where $T_{0}=0.9975$, $\rho _{v}=0.955$ and $\rho _{l}=1.045$. Parameters are $
\tau =10^{-4}$, $\Delta t=5\times 10^{-5}$, $\Delta x=\Delta y=2\times
10^{-2}$, $K=10^{-5}$, $H=0$, $\zeta =0$, $q=-0.004$, $b=2$. Here the initial
temperature is $0.9975$. It will be dropped by a small value when the
equilibrium state has been achieved. Figure 1(a) shows the liquid-vapor coexistence curve obtained from the LB
simulations at various temperatures, compared to the theoretical predictions
from Maxwell construction. The two sets of results have a satisfying
agreement. Moreover, the lowest temperature that the D2V16-FFT-TLB model can
simulate is about $0.8$, while for the D2V33-FFT-TLB model, it is about $0.85$. So the
former is stuitable for a wider temperature range and a wider range of density ratio.

\begin{figure}[tbp]
\center {\epsfig{file=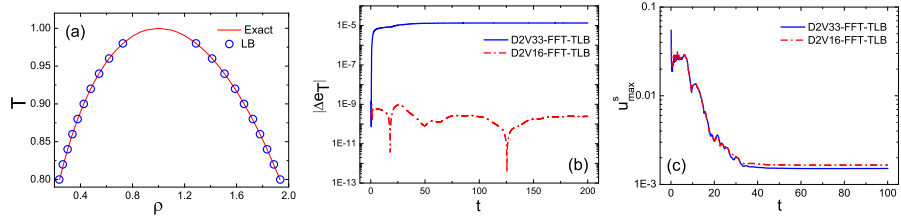,bbllx=108pt,bblly=389pt,bburx=566pt,bbury=513pt,
width=0.94\textwidth,clip=}}
\caption{(Color online) (a) Comparisons of the liquid-vapor coexistence curve obtained from LB
simulations and Maxwell construction. (b) Variations of the total energy $|\Delta
e_{T}(t)|=|e_{T}(t)-e_{T}(0)|$ for the phase separating procedure described
in Fig.2. (c) Evolution of the maximum spurious velocity $u_{\max}^{s}$ during the large deformation procedure displayed in Fig. 3.}
\end{figure}

\subsection{Phase separation after quench}

\begin{figure}[tbp]
\center {
\epsfig{file=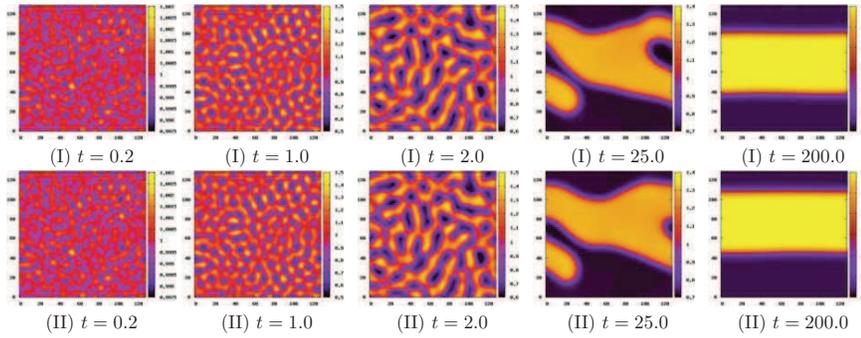,bbllx=0pt,bblly=0pt,bburx=474pt,bbury=186pt,
width=0.94\textwidth,clip=}}
\caption{(Color online)Density patterns at representative times during phase separation,
where (I) stands for results obtained from the D2V16-FFT-TLB model and (II)
from the D2V33-FFT-TLB model. }
\end{figure}

Here the thermal phase separation after quench is employed to validate if
the new model can recover the accurate physical scenario. Simulations
are performed on a $N_{x}\times N_{y}=128\times 128$ lattice nodes. The initial conditions are
set as $(\rho ,T,u_{x},u_{y})=(1.0+\Delta ,0.9,0.0,0.0)$, where $\Delta $ is
a random density noise with an amplitude of $0.001$. Parameters are set to
be $\tau =10^{-4}$, $\Delta t=10^{-5}$ for D2V33-FFT-TLB model and $10^{-4}$
for D2V16-FFT-TLB model, $K=5\times 10^{-6} $, $\Delta x=\Delta y=1/256$.
The FTT scheme with 8th order in precision and the
1st Euler scheme are used for the D2V33-FFT-TLB model, while the FFT scheme
with 16th order in precision and the 2nd RK scheme are for the D2V16-FFT-TLB
model.

Typical snapshots of configurations are plotted in Fig. 2.
After quench the fluid begins to separate spontaneously into small regions with higher
and lower densities. The density difference increases with time.
Under the actions of surface tension and viscous force,
small domains coalesce to each other and larger ones appear at about $t=1.0$. From
patterns at $t=1.0$ and $t=2.0$, the liquid and vapor domains evolve
in an equal way, leading to an interwoven bicontinuous pattern. The growth
of domains continues at $t=25.0$, and eventually at $t=200.0$, the system
reaches a completely separated state with least surface.

Compared to the D2V33-FFT-TLB model, the D2V16-FFT-TLB model recovers nearly
the same results, particularly at former times. With the accumulation of
physical and numerical differences between the two models, the patterns
present larger differences. For example, at $t=25.0$, the vapor phase at the
right boundary calculated from the D2V16-FFT-TLB model is much larger than
that calculated from the D2V33-FFT-TLB model. And at $t=200.0$, the liquid
band simulated by the D2V16-FFT-TLB model is located lower than the one
mimicked by the later. However, if we focus on the evolution mechanism of
the system, rather than the morphologies at specific times, the differences
can be neglected. Furthermore, physically, due to the utilization of the DVM
with the least discrete velocities and, numerically, due to application of
higher-order RK scheme for temporal derivative, the D2V16-FFT-TLB model is
about 10 times faster than the D2V33-FFT-TLB model. This is of essential
importance for studying a phase-separating system, because the enhancement
in computational efficiency means the decrease in simulation time we needed,
the increase in lattice nodes we employed and the feasibility of the
simulation on a personal computer.

Figure 1(b) depicts variations of the total energy $|\Delta e_{T}|$ for the
process shown in Fig. 2. To be seen is that the
D2V16-FFT-TLB model not only owns higher efficiency but also higher
accuracy and better energy conservation in simulations.
When the new model is used, $|\Delta
e_{T}| $ remarkably reduces from $10^{-5}$ to $10^{-9}$.

\subsection{Large deformation of a rectangular droplet}

\begin{figure}[tbp]
\center {
\epsfig{file=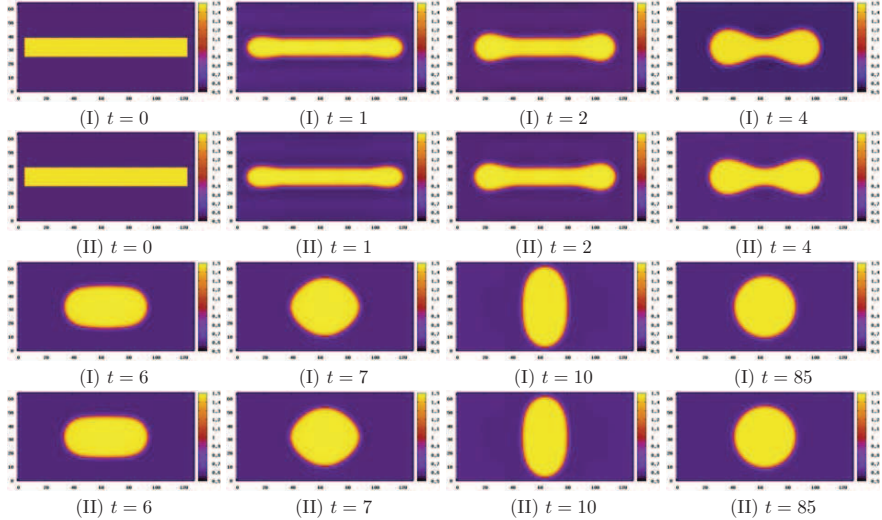,bbllx=0pt,bblly=0pt,bburx=474pt,bbury=280pt,
width=0.94\textwidth,clip=}}
\caption{(Color online) Snapshots for a large deformation of a rectangular
droplet at typical times, where (I) stands for results obtained from
the D2V16-FFT-TLB model and (II) from the D2V33-FFT-TLB model. }
\end{figure}
To further assess if the new model is appropriate for simulating large
deformation problem and examine its ability for refraining spurious
velocity, a large deformation problem under surface tension is studied.
Initially a rectangular droplet with length $118$ and height $14$ is placed
at the center of the whole domain, where the lattice unit is used. The liquid and vapor densities are $\rho
_{l}=1.462$ and $\rho _{v}=0.579$, respectively. Throughout the simulation,
the temperature $T$ is fixed to be $0.95$. Parameter are $\Delta x=\Delta
y=0.01$, $\tau =5\Delta t=5\times 10^{-4}$, $K=10^{-5}$ and $q=-0.05$. For
both models, the FFT scheme with 16 order in accuracy and the 2nd RK are employed.
Figure 3 illustrates snapshots of the droplet shapes at various times. When the
simulation starts, the droplet will firstly transform its configuration at
the left and right sides, where the surface tension reaches its maximum due
to the largest curvature. As time evolves, the droplet gradually shrinks to
a dumbbell shape for minimizing the net interfacial energy. At $t=6.0$, it
becomes approximately an ellipse. After that, it oscillates through changing
the length of the major axis and minor axis. Due to the effects of surface
tension and viscosity, the deformed droplet relaxes to a circular
equilibrium shape at about $t=85.0$. The simulation results obtained from
the two models are consistent with each other, and are qualitatively
agreement with Lamb's theoretical analysis\cite{Lamb} and other researchers'
results\cite{JCP-2006}.

Evolution of the maximum spurious velocity $u_{\max}^{s}$ during the large
deformation procedure is illustrated in Fig. 1(c). When $t>40.0$,
the system approximately achieve its equilibrium and $u_{\max}^{s}$ almost
remains unchanged. With the identical numerical algorithms, the new model
has the similar ability to refrain $u_{\max}^{s}$.

\section{Conclusions}

In the paper, a highly efficient LB model for thermal liquid-vapor system is
proposed. Due to the applications of DVM with less
discrete velocities and higher-order
schemes, the new model is more efficient, has more potential to keep
conservation of the total energy and refrain spurious velocity near the
liquid-vapor interface. Therefore, it is more suitable for studying
thermohydrodynamics of multiphase flows. In future studies, by inserting
more realistic EOS, we will increase the density ratio that the new model
can simualte, and will investigate the hydrodynamic instabilities which are highly
interested in science and engineering and where various nonequilibrium effects are pronounced.


\section*{Acknowledgements}
This work is partly supported by
National Natural Science Foundation of China (under Grant Nos. 11071024,
11202003 and 11203001), Foundation of State Key Laboratory of Explosion
Science and Technology (under Grant No. KFJJ14-1M), Natural Science
Foundation of Hebei Province (under Grant No. A2013409003),
Foundation for Outstanding Young Scholars of Hebei Province,
Excellent Youth Foundation of Hebei Educational committee (under Grant No. YQ2013013) and
Technology Support Program of LangFang (under Grant Nos. 2011011023,
2012011021/26/30).

\end{document}